\documentclass[onecollarge,natbib]{svjour2}
\bibpunct{[}{]}{;}{n}{}{,} 
\smartqed  
\usepackage{graphicx}
\newcommand{\cd}{\makebox[0.08cm]{$\cdot$}}
\journalname{Few Body Systems}
\begin{document}

\title{Present status of
the Bethe-Salpeter approach in Minkowski space
}


\author{V.A.~Karmanov        
}


\institute{V.A.~Karmanov \at
              Lebedev Physical Institute, Leninsky Prospect 53, 119991 Moscow, Russia \\
              \email{karmanov@sci.lebedev.ru}           
}

\date{Received: date / Accepted: date}

\maketitle

\begin{abstract}
Recently developed methods allowing to find the solutions of the Bethe-Salpeter equations in Minkowski space, 
both for the bound and scattering states, are reviewed. For the bound states, one obtains the bound state mass and the corresponding BS amplitude. For the scattering states, 
the phase shifts (complex above the meson creation threshold) and the half-off-shell BS amplitude are found. Using these solutions, the elastic and transition electromagnetic form factors are calculated.
\end{abstract}

\section{Introduction}
\label{intro}



Bethe-Salpeter (BS) approach \cite{bs} is a powerful tool in the theory of relativistic few-body systems.
During the last few years there has been a renaissance of this approach in its original Minkowski space formulation.
It is caused by the progress in finding new methods which allow to overcome the difficulties resulting 
from the singularities of the BS equation and to obtain the off-mass shell solution in the Minkowski space both for the bound and scattering states. 
The on-shell solution for the scattering states, found  few decades ago \cite{tjon,Schwartz_Morris_Haymaker},  allows to compute the nucleon-nucleon (NN) scattering phase shifts. 
The off-shell solution for the bound \cite{bs1,bs2} and scattering states \cite{KC_LCM_Cracovie_2012,kc_FB20,CK_Baldin_2012,CK_PLB_2013} is much more recent
and allows to properly calculate the electromagnetic  elastic from factor  \cite{ckm_epja,ck-trento}
as well as the  transition one \cite{ck-transit} describing the electro-disintegration of  a bound system (e.g. the deuteron). 

For that purpose one needs the Minkowski space solution since  the Euclidean  one cannot be used: the Wick rotation can be done in the BS equation itself, 
but not in the integral expression giving the form factors \cite{ck-trento}. In this integral, the rotated contour crosses singularities
which prevent from the Wick rotation. 
As demonstrated explicitly in \cite{ck-trento} (sect. 6) in a simplest example  when the BS vertices are constant, 
still one singularity is on the way of rotating contour. In a more realistic case, when interaction is determined by the one-boson exchange kernel, the BS amplitude takes implicitly into account 
all the numbers of intermediate bosons (the thresholds with the virtual  $1, 2, n,\ldots$ bosons) that results in infinite number of singularities.

Analogously to finding the nonrelativistic NN potential starting with the phase shifts, one can also fit, in a self-consistent relativistic version, the BS kernel and then use it in further calculations. 
This full BS program has been achieved for a  NN system interacting via separable kernel \cite{burov} 
since in this case, the singular integrals -- both in the equation and in the form factors -- are performed analytically.
However it is not yet fully realized  for a field-theoretically inspired description. 

For the one-boson exchange or more complicated kernel the analytical treatment of singularities is also possible when the bound state BS solution is found in the form of the Nakanishi integral representation \cite{nak63,nak71}.  In this case one can  express the form factor \cite{ckm_epja} through the non-singular Nakanishi weight function.
For the scattering states the development of this method (based on the Nakanishi representation  \cite{nak63,nak71}) is in progress \cite{fsv-2012} but  the numerical calculations are not yet presented. 
On the other hand, we have recently developed another approach \cite{KC_LCM_Cracovie_2012,kc_FB20,CK_Baldin_2012,CK_PLB_2013}
based on the direct treatment of the singularities (without using the Nakanishi representation) which also  allows to find both the bound and scattering state solutions and calculate e.m. form factors.


The methods based on the Nakanishi representation and the corresponding results, both for spineless particles and fermions, are described in detail in the recent review paper \cite{ck-trento}. Therefore we will only sketch them here in sect. \ref{sect_nak}. Whereas, the solutions found using the direct treatment of the singularities, mainly for the scattering states,  will be described in more detail in sect \ref{sect_scat}. With these solutions, the transition form factor is calculated in sect. \ref{transit-ff}. The concluding remarks are given in sect. \ref{concl}.

\section{Method based on the Nakanishi representation}
\label{sect_nak}
\begin{figure}
\centering
\includegraphics{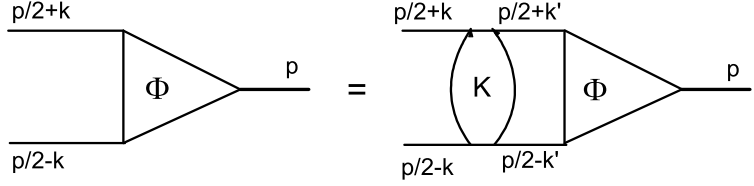}
\caption{Graphical representation of the BS equation for the bound state.}
\label{fig1}       
\end{figure}
For a bound state of total momentum $p$ and in case of equal mass spinless
particles with zero angular momentum, the BS equation for the BS amplitude $\Phi$ is shown graphically in figure \ref{fig1}. It reads
\begin{equation}\label{bs}
\Phi(k,p)=\frac{i^2}{\left[(\frac{p}{2}+k)^2-m^2+i\epsilon\right]
\left[(\frac{p}{2}-k)^2-m^2+i\epsilon\right]} \int
\frac{d^4k'}{(2\pi)^4}iK(k,k',p)\Phi(k',p),
\end{equation}
where $iK$ is the interaction kernel, $m$ is
the mass of the constituents and $p=p_1+p_2$, $k=(p_1-p_2)/2$
are their total and relative four-momenta ($p_{1,2}$ are the constituent momenta).
Note that 
$\Phi(k,p)$ includes, by definition, the external propagators.
We will denote by $M=\sqrt{p^2}$ the total mass of the bound state. 
Being scalar, the BS amplitude $\Phi(k,p)$ depends on the scalar products
$k^2$ and $p\cd k$:  $\Phi(k,p)=\Phi(k^2, p\cd k)$. In the c.m.-frame $\vec{p}=0$ we get $p\cd k=Mk_0$. Since $k^2=k_0^2-\vec{k}^2$, 
the BS amplitude in this frame (for S-wave) depends on two variables $|\vec{k}|,k_0$: $\Phi=\Phi(|\vec{k}|,k_0)$. 

Then one can make the Wick rotation \cite{WICK_54}, i.e. to replace the integration over $-\infty < k_0 < \infty$  in the integration volume $d^4k=dk_0d^3k$ by $-i\infty < k_0 < i\infty$. It is equivalent to the replacemeant of variable $k_0=ik_4$ 
where $k_4$ is real and varies in the limits $-\infty < k_4 < \infty$. In this way we obtain the non-singular Euclidean BS equation:
\begin{equation}\label{bsE}
\left[\left(m^2-\frac{M^2}{4}+\vec{k}^2+k_4^2\right)^2+M^2k_4^2\right]\Phi_E(\vec{k},k_4)=
 \int\frac{d^3k'dk_4}{(2\pi)^4}K_E(k,k')\Phi_E(\vec{k'},k_4)
\end{equation}
where $\Phi_E(\vec{k},k_4)=\Phi(\vec{k},ik_4)$ and also $K_E(\vec{k},k_4;\vec{k'},k'_4)=K(\vec{k},ik_4;\vec{k'},ik'_4)$. Indeed, the factor in the square brackets in the l.h.-side of (\ref{bsE}) never crosses zero.
Take, for example, the Minkowski one-boson exchange kernel
\begin{equation}\label{obe}
K(k,k',p)=\frac{-16\pi m^2\alpha}{(k_0-k'_0)^2 -(\vec{k}-\vec{k'})2-\mu^2+i\epsilon},
\end{equation}
where $\alpha=g^2/(16\pi m^2)$  is the dimensionless coupling constant  of the Yukawa potential $V(r)=-\alpha\exp(-\mu r)/r$.
It is singular (if $i\epsilon =0$, the denominator crosses zero in the integration domain). After Wick rotation it turns into the non-singular Euclidean kernel
\begin{equation}\label{obeE}
K_E=\frac{16\pi m^2\alpha}{(k_4-k'_4)^2 +(\vec{k}-\vec{k'})2+\mu^2}
\end{equation}
(the denominator is always positive and never crossed zero).
The numerical resolution of eq. (\ref{bsE}) is a trivial task. It gives exactly the same mass $M$ as defined by the Minkowski BS equation (\ref{bs}),
but the amplitude $\Phi_E(\vec{k},k_4)$ differs from one entering the equation (\ref{bs}).

In refs. \cite{KW, KSW, bs1,bs2,sauli} the Minkowski BS amplitude was found in the form of the Nakanishi integral representation
\cite{nak63,nak71}:
\begin{equation}\label{bsint}
\Phi(k,p)=\int_{-1}^1\mbox{d}z'\int_0^{\infty}\mbox{d}\gamma'
\frac{g(\gamma',z')}{\left[  k^2+p\cdot k\; z' +\frac{1}{4}M^2-m^2
-  \gamma' + i\epsilon\right]^3}.
\end{equation}
In this representation  the dependence on the two
scalar arguments $k^2$ and $p\cd k$  of the BS amplitude is made
explicit by the integrand denominator and the Nakanishi  weight
function $g(\gamma,z)$ is non-singular. 

By inserting
the amplitude (\ref{bsint}) into the BS equation one finds an
integral equation, still containing  singularities, however for non-singular $g$.
To eliminate these singularities from equation, we apply to both sides of equation  an
integral transform -- light-front projection  \cite{bs1}. 
It consists in the
replacing $k\to k+\frac{\omega}{\omega \cd p}\,\beta$ where
$\omega$ is a light-cone four-vector $\omega^2=0$, and integrating
over $\beta$ in infinite limits. We obtain in this way,  a
non-singular equation for the non-singular $g(\gamma,z)$:
\begin{equation} \label{bsnew}
\int_0^{\infty}\frac{g(\gamma',z)d\gamma'}{\Bigl[\gamma'+\gamma
+ m^2-\frac{1}{4}(1-z^2)M^2\Bigr]^2} =
\int_0^{\infty}d\gamma'\int_{-1}^{1}dz'\;V(\gamma,z;\gamma',z')
g(\gamma',z'),
\end{equation}
The kernel $V(\gamma,z;\gamma',z')$ here is expressed through an integral  \cite{bs1} from the kernel  $K(k,k',p)$ entering
in the initial BS equation (\ref{bs}). The total mass
$M$ of the system (to be found) appears on both sides of equation (\ref{bsnew}).
After
solving eq. (\ref{bsnew}) and substituting the solution in eq. (\ref{bsint}), one finds the
BS amplitude in Minkowski space which  can be used to calculate the e.m. form factor.

\begin{figure}
\centering
\includegraphics[width=0.4\textwidth]{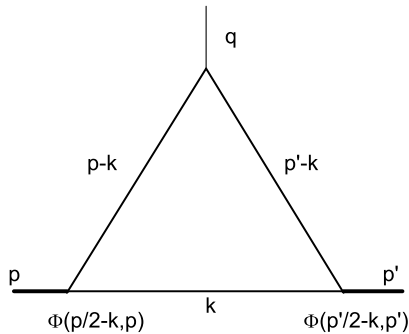}
\caption{E.m. vertex in terms of the BS amplitude.}
\label{fig2}       
\end{figure}

The equation (\ref{bsnew}) was solved for the ladder kernel in \cite{bs1} and for the ladder + cross-ladder kernel in \cite{bs2}. 
The bound state mass $M$ coincides with the one found from the BS equation in the Euclidean space.
After substituting the Minkowski space solution, found in the form (\ref{bsint}), in the expression for the e.m. form factor (shown graphically in figure \ref{fig2}):
\begin{equation}\label{ff}
(p+p')^\mu F(Q^2) =-i\int \frac{d^4k}{(2\pi)^4}(p+p'-2k)^\mu \; (m^2-k^2)\Phi \left(\frac{p}{2}-k,p\right)\Phi  \left(\frac{p'}{2}-k,p'\right)
\end{equation}
and using the Feynman parametrization,
the 4D integral over four-momentum $k$ is calculated analytically and the form factor is expressed through the non-singular integral containing $g$ quadratically \cite{ckm_epja}.

The same approach was applied in \cite{ck-ferm} for solving the BS equation for the two-fermion system interacting by the scalar, pseudo scalar and massless vector exchanges (one-boson exchange kernel). It turned out that, in contrast to the spineless case, the light-front projection does not eliminate completely the 
singularities of the kernel in the equation for the Nakanishi function.   These singularities are integrable numerically. They do
not prevent from finding numerical solution, but they reduce its precision. This difficulty can be avoided by a proper regularization of the BS
equation \cite{ck-ferm}.
After taking the light-front projection, we get system of equations similar to (\ref{bsnew}) for the set of the Nakanishi functions $g_i(\gamma,z)$, corresponding to the spin components of the two-fermion system.
The bound state masses again coincide with ones found from the Euclidean BS equation \cite{dorkin}.
The approach based on the Nakanishi representation  was used in \cite{sauli},  under some simplifying
ansatz,  to solve the BS equation for a quark-antiquark system.

\section{Scattering states}
\label{sect_scat}
\begin{figure}[h!]
\centering
\includegraphics[width=11.cm]{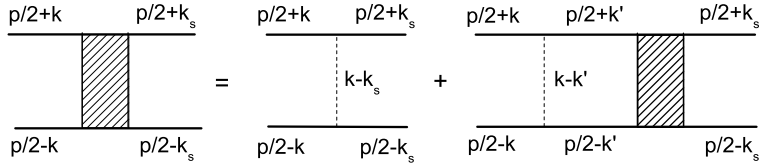}
\caption{Bethe-Salpeter equation for a scattering state.}\label{bs_eq}
\end{figure}
The inhomogeneous scattering state BS equation is graphically represented in figure \ref{bs_eq}.  In Minkowski space it reads:
\begin{equation}\label{BSE}
F(k,k_s;p)=K(k,k_s;p)- i\int\frac{d^4k'}{(2\pi)^4}
\frac{K(k,k';P) F(k',k_s;p)}
{\left[\left(\frac{p}{2}+k'\right)^2-m^2+i\epsilon\right]
\left[\left(\frac{p}{2}-k'\right)^2-m^2+i\epsilon\right]}
\end{equation}
We denote by $k$ the relative four-momentum -- variable of the equation, $k_s$  the
scattering (incident) relative momentum and $p$  the total momentum of the state with $p^2=M^2$,  the squared total invariant mass of the system.
We will  consider  
the spinless particles ($m$=1) interacting by the one-boson exchange kernel (\ref{obe}).

The half-off-shell scattering amplitude $F(k,k_s;p)$ satisfying eq. (\ref{BSE}), in contrast to the bound state BS amplitude from eq. (\ref{bs}), does not include in itself the propagators of external particles. It depends on the three four-momenta $k,k_s,p$.  For a given incident  momentum $\vec{k}_s$ and written in the center of mass frame $\vec{p}=0$, $p_0=M=2\varepsilon_{k_s} =2\sqrt{m^2+k_s^2}$, $k_{0s}=0$,
$F$  depends on three variables  $|\vec{k}|$, $k_0$ and $z=\cos(\vec{k},\vec{k}_s)$.
It will be 
denoted  by $F(k_0,k,z;k_s)$, setting abusively $k=|\vec{k}|$, $k_s=|\vec{k_s}|$.
The module of incident momentum $k_s$ plays role of parameter (like the bound state mass $M$ in the bound state equation (\ref{bs})),
whereas, in contrast to the bound state case, $F$ depends also on the extra variable --  cosine $z$ of the scattering angle.

The amplitude we consider here is a particular case of the so called full off-shell amplitude\\ $F_0(k_0,k,z;k_{0s},k_s;M)$.  
The latter, in addition to the variables $k_0,k$ depends also on the off-shell independent variables $k_{0s},k_s$, 
now with $k_{0s}\neq 0$ and $k_{0s}\neq \varepsilon_{k_s}$. The total mass $M=\sqrt{s}$ is neither equal to $2\varepsilon_{k_s}$ nor related to $k_{0s}$. 
By "off-shell amplitude" we will hereafter mean half-off-shell amplitude. The method we have developed  is also applicable to the full off-shell amplitude, 
though dependence of the latter on two extra variables $k_{0s},k_s$ requires much more extensive numerical calculations.

Depending on three variables, the scattering amplitude $F$ is represented now through the three-parameter Nakanishi integral (comparer with eq. (58) from \cite{fsv-2012}):
\begin{equation}\label{nakanishi-scat}
F(k,k_s;p)=\int_{-1}^1 dz' \int_{-1}^1 dz''\int_{-\infty}^{\infty}d\gamma' \frac{g(z',z'',\gamma')}
{ [\gamma'+m^2-\frac{1}{4}M^2-k^2-p\cd k\,z''-2k\cd k_s\,z'-i\epsilon]}
\end{equation}
Note that the amplitude $F(k,k_s;p)$ is  not decomposed in the partial waves.
The equation for the function $g(z',z'',\gamma')$ was derived in \cite{fsv-2012} but not yet solved numerically.

On the other hand, we have developed recently \cite{KC_LCM_Cracovie_2012,kc_FB20,CK_Baldin_2012,CK_PLB_2013} another method which allows to avoid the Nakanishi representation and 
find the BS amplitudes in Minkowski space both for the bound and scattering states. 

We use the partial wave decomposition following its definition from \cite{IZ}:
\begin{equation}\label{partw} 
F(k_0,k,z;k_s)=16\pi\sum_{l=0}^{\infty}(2l+1) F_l(k_0,k)P_l(z)
 \end{equation}
For illustration, we will consider the S-wave only, through there is no any problem to take arbitrary partial wave $l$.
In the given normalization, the on-shell amplitude $F^{on}_0(k_s)\equiv F_0(k_0=0,k=k_s)$ determines the phase shift according to:
\begin{equation}\label{delta}
\delta_0(k_s)=\frac{1}{2i}\log\Bigl(1+\frac{2i k_s } {\varepsilon_{k_s}} F^{on}_0(k_s)\Bigr)
\end{equation}

Several steps must be accomplished  before obtaining a
solvable equation for $F_0$ which takes into account  the singularities of the BS equation. The main one is the transformation of the initial BS equation  (\ref{BSE}) into the form in which the singularities of the constituent propagators are eliminated. 
The propagators in the  r.h.-side of  (\ref{BSE}) have two poles, each of them represented as sum of principal value and  $\delta$-function.
Their product gives rise to terms having respectively 0, 1 and 2 $\delta$'s. After partial wave decomposition, the 4D equation (\ref{BSE}) is reduced into a 2D one.
Integrating analytically  over $k'_0$ the $\delta$ contributions and eliminating the
principal values singularities by the standard subtractions (see eq. (\ref{PV}) below), we derive the following S-wave equation \cite{CK_PLB_2013}:
\begin{eqnarray}
&&F_0(k_0,k)  =  F^B_0(k_0,k)    +  \frac{i\pi^2 k_s}{8\varepsilon_{k_s}} W_0^S(k_0,k,0,k_s) F_0(0,k_s)
 \cr
                 &+& \frac{\pi}{2M} \int_0^{\infty}  \frac{dk'}{ \varepsilon_{k'} ( 2\varepsilon_{k'}-M) }    \left[{k'}^2 W_0^S(k_0,k, a_-,k') F_0(| a_- |,k')
     -\frac{2 {k_s}^2\varepsilon_{k'}}{\varepsilon_{k'} +\varepsilon_{k_s}}W_0^S(k_0,k,0,k_s) F_0(0,k_s)\right]
\cr
                    &-&    \frac{\pi}{2M} \int_0^{\infty} \frac{{k'}^2 dk'}{\varepsilon_{k'} (2\varepsilon_{k'}+M) }      W_0^S(k_0,k, a_+,k') F_0(a_+,k' )   \cr
                 &+&  \frac{i}{2M}  \int_0^{\infty}  \frac{{k'}^2dk' }{\varepsilon_{k'}}   \int_0^{\infty} dk'_0 \left[ \frac{ W^S_{0}(k_0,k,k'_0,k') F_0(k'_0,k')  - W^S_{0}(k_0,k,a_-,k')  F_0(|a_- |,k')}{  {k'}_0^2-a_-^2 }\right]  \cr
                &-&    \frac{i}{2M}  \int_0^{\infty} \frac{{k'}^2dk'} {\varepsilon_{k'}}     \int_0^{\infty} dk'_0 \left[ \frac{ W^S_{0}(k_0,k,k'_0,k') F_0(k'_0,k')  - W^S_{0}(k_0,k,a_+,k') F_0(a_+,k')}  {{k'}_0^2-a_+^2 } \right]
                   \label{Eq_F_sym}
\end{eqnarray}
where $a_{\mp} =\varepsilon_{k'} \mp \varepsilon_{k_s}$ and $W_0^S$ is the S-wave kernel -- suitably symmetrized on $k'_0$ variable to restrict its integration domain to $[0,\infty]$ --
is given by
\[ W_0^S(k_0,k,k'_0,k')=W_0(k_0,k,k'_0,k')+W_0(k_0,k,-k'_0,k')\]
with:
\begin{equation}\label{W0}
W_0(k_0,k,k'_0,k') = -\frac{\alpha m^2}{\pi kk'} \left\{  \frac{1}{\pi}  \log \left| \frac{(\eta+1)}{(\eta-1)} \right| - i I(\eta)   \right\},
\quad
I(\eta)=\left\{
\begin{array}{lcrcl}
1  & {\rm if} & \mid\eta\mid &\leq& 1 \cr
0  & {\rm if} & \mid\eta\mid &> & 1
\end{array}\right.
\end{equation}
and
\[  \eta =   {1\over2kk'} \left[ (k_0  - k'_0)^2 - k^2 - {k'}^2 -\mu^2  \right]  \]

The inhomogeneous (Born)  term $F^B_0$ reads:
\[ F^B_0(k_0,k)={\pi^2\over 4}W_0(k_0,k,0,k_s)\]

The origin of the different terms appearing in (\ref{Eq_F_sym})
is the following. The non-integral term in the first line
follows from the integrated (2D) product of the two $\delta$-functions   mentioned above.
The one-dimensional integrals -- second and third lines  --  result from  one $\delta$-function terms, after integration over $k'_0$. 
The last two lines come from the principal values (PV) alone. 
The differences appearing in the squared brackets correspond to removing the pole singularities at $2\epsilon_{k'}=M$   (second line) and $k'_0=a_{\mp}$
(forth and fifth lines) according to the well known subtraction technique eliminating singularity:
\begin{equation}\label{PV}
PV\int_0^{\infty} \frac{f(x')}{{x'}^2-a^2}\;dx'=\int_0^{\infty} \frac{f(x')-f(a)}{{x'}^2-a^2}\;dx'
\end{equation}
In l.h.-side of (\ref{PV})  the integrand  at $x'=a$ is singular that complicates the numerical calculation of integral, whereas r.h.-side does not contain this singularity.

We also treat analytically the singularities of the kernel (\ref{obe}), which after the  partial wave decomposition are the logarithmic ones (see eq. (\ref{W0})),
the singularities of the Born term and of  the amplitude $F_0$ itself. 
We obtain in this way a non-singular equation which we solve by standard methods. 

Our first check was to solve the bound state problem by dropping the inhomogeneous term in (\ref{BSE}) and setting $M=2m-B$.
The binding energy $B$  thus obtained,  coincides within four-digit accuracy, with the one calculated, by other method, in our previous work \cite{bs1}.
Other, much more sophisticated tests \cite{CK_PLB_2013} are also satisfied. We also reproduce, with a reasonable accuracy, the phase shifts found in \cite{tjon}.

The BS relativistic formalism accounts naturally for  the meson creation in the scattering process, when the available  kinetic energy allows it.
Below the first inelastic threshold,  $k^{(1)}_s=\sqrt{m\mu+\mu^2/4}$, the phase shifts given by eq. (\ref{delta}) are real.
This unitarity condition is not automatically fulfilled in our approach, but appears as a consequence of handling the correct solution $F^{on}_0(k_s)$
and provides a stringent test of the numerical method. It is violated for any minor distortion of the equation (\ref{Eq_F_sym}) or of the solution $F_0(k_0,k)$.
Above  $k^{(1)}_s$, the phase shift obtains an imaginary part which behaves numerically in the threshold vicinity like
$ {\rm Im}(\delta_0)\sim (k_s-k^{(1)}_s)^2$,
as expected.
Higher inelasticity thresholds, corresponding to creation of 2, 3, etc. intermediate mesons, are also  taken into account in our calculations.

The S-wave off-shell scattering amplitude $F_0(k_0,k)$  in Minkowski space  was thus  safely computed and the phase shifts, presented below, are extracted according to (\ref{delta}).
Some details are given in \cite{CK_PLB_2013}. Full details will be presented in a paper which is under preparation.

\subsection{Numerical results}

\begin{figure}[h]
\centering
\includegraphics[width=7.cm]{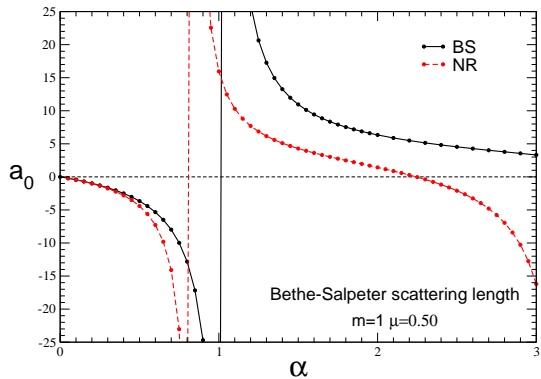}
\caption{BS scattering length $a_0$ versus the coupling constant  $\alpha$ (solid), compared to the non-relativistic results (dashed) for $\mu=0.5$.}\label{fig_a0}
\end{figure}

The low energy parameters were computed
and found to be consistent with a quadratic fit to the effective range function $k\cot\delta(k)= -\frac{1}{a_0} + \frac{1}{2} r_0 k^2 $ .
The BS scattering length  $a_0$ as a function of the coupling constant $\alpha$ is given in figure \ref{fig_a0} for  $\mu=0.50$.
It is compared to the non-relativistic (NR)  values provided by the Schr\"odinger equation with the Yukawa potential.
The singularities correspond to appearance of the first bound state at $\alpha_0=1.02$ for BS and  $\alpha_0=0.840$ for NR.
As one can see, the differences between a relativistic and a non-relativistic treatments of the same problem
are not of kinematical origin since even for  processes involving zero energy they  can be substantially large, especially in presence of bound state.

\begin{figure}[ht!]
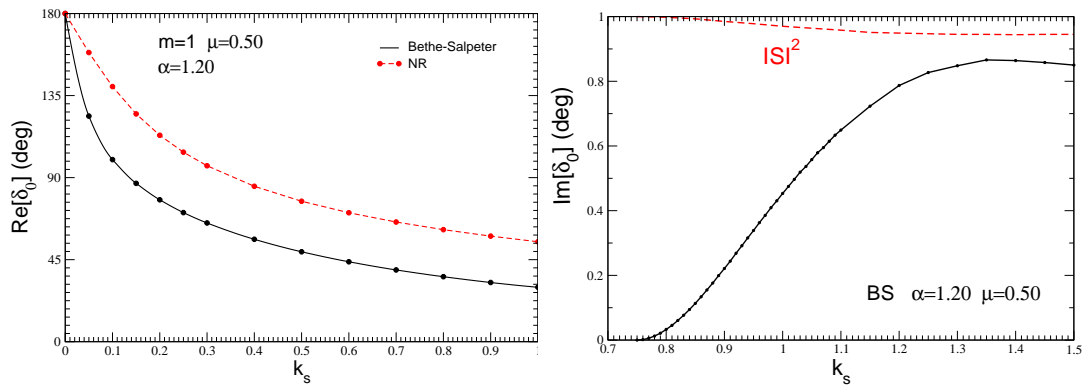

\vspace{0.5cm}
\begin{center}
\includegraphics[width=7cm]{Mod_Phaseshifts_alpha_1.200_0.50.eps}
\includegraphics[width=7cm]{Mod_Inel_alpha_1.20_mu_0.50.eps}
\end{center}
\caption{Real (left panel) and imaginary (right panel)  phase shift (degrees) for $\alpha=1.2$ and $\mu=0.50$ calculated  via BS equation in the form (\protect{\ref{Eq_F_sym}}) (solid)  compared to the non-relativistic results (dashed).}\label{fig2a}
\end{figure}

Figure \ref{fig2a}  (left panel) shows the real  part of the phase  shifts calculated with BS  (solid line)  and NR (dashed line) equations and with the same parameters than in figure \ref{fig_a0}.
For this value of $\alpha$ there exists a bound state and, according to the Levinson theorem, the phase shift starts at 180$^\circ$.
One can see that the difference between relativistic and non-relativistic results is considerable even for relatively small incident momentum.
The right panel  shows the imaginary part of the phase shift. It appears  starting from the first inelastic meson-production threshold $k^{(1)}_s=0.75$ and displays the expected quadratic behavior vs. $k_s$.
Simultaneously the modulus squared of the S-matrix (displayed in dashed line) starts to differ from unity.
The value $Im[\delta]$ in this figure contains  the contributions of the second $k^{(2)}_s=1.118$ and third  $k^{(3)}_s=1.435$ meson creation thresholds as well.

\begin{figure}[ht!]
\begin{center}
\includegraphics[width=7cm]{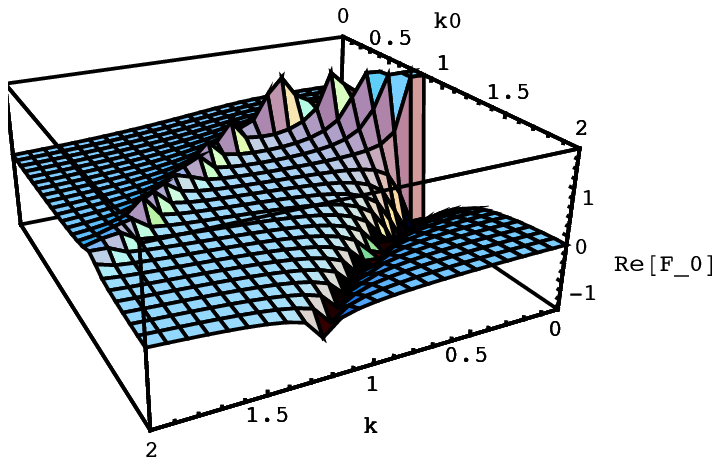}
\includegraphics[width=7cm]{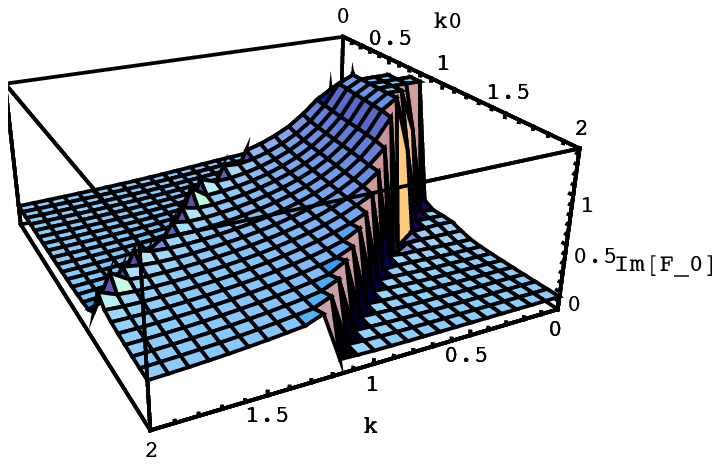}
\end{center}
\caption{Real  (left panel) and imaginary (right panel) parts of the off-shell amplitude $F_0(k_0,k)$ for $\alpha=0.5$, $k_s=\mu=0.5$.} \label{fig3}
\end{figure}
Finally, we display in figure \ref{fig3} the real (left panel) and imaginary (right panel) parts of the off-shell scattering amplitude $F_0(k_0,k)$
as a function of  $k_0$ and $k$ calculated for $\alpha=0.5$, $k_s=\mu=0.5$.
It shows a non trivial structure with  a ridge and a gap in the real part and a plato in the imaginary part resulting from the singularities of the inhomogeneous term.
Its on-shell value $F_0^{on}=F_0(k_0=0,k=k_s=0.5)=0.753+i0.292$,
determining  the phase shift $\delta=21.2^{\circ}$, corresponds to a single point on theses two surfaces accessible by the previous methods \cite{tjon,Schwartz_Morris_Haymaker}.
Our calculation, shown in figure \ref{fig3}, provides   the full amplitude $F_0(k_0,k)$ in  a two-dimensional domain.

Computing this quantity, and  related on-shell observables, was the main result of the works \cite{KC_LCM_Cracovie_2012,kc_FB20,CK_Baldin_2012,CK_PLB_2013}.
Together with the bound state solution in Minkowski space \cite{bs1,bs2},  they pave the way for a consistent relativistic description of composite systems in the framework of  BS equation.

\section{Transition form factor}
\label{transit-ff}

\begin{figure}[h!]
\begin{center}
\includegraphics[width=0.4\textwidth]{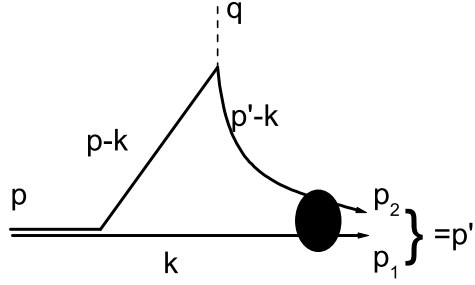}
\caption{Feynman diagram for the EM transition form factor.}\label{triangle}
\end{center}
\end{figure}

As soon as we know the Minkowski BS amplitude for the bound state, we can calculate, by eq. (\ref{ff}), the elastic e.m. form factor.
Knowing both amplitudes: for bound and scattering states, we can calculate the transition form factor corresponding to electrodisintegration
of the bound system.  This calculation \cite{ck-transit} will be briefly explained below.

We start with the expression for the transition current, represented graphically  in figure \ref{triangle} (the plane wave contribution is not  considered):
\begin{equation}\label{ffin0}
\tilde{J}_{\mu}=i\int \frac{d^4k}{(2\pi)^4}\,
\frac{(p_{\mu}+p'_{\mu}-2k_{\mu})\Gamma_i \left(\frac{1}{2}p -k,p\right)\Gamma_f
\left(\frac{1}{2}p'-k,p'\right)} {(k^2-m^2+i\epsilon)[(p-k)^2-m^2+i\epsilon]
[(p'-k)^2-m^2+i\epsilon]}.
\end{equation}
Here $\Gamma_i$ is  the initial (bound state) vertex function, defined by (\ref{BSE}) without inhomogeneous term,  $\Gamma_f=16\pi F_0$ in the final one (the factor $16\pi$ is  due to eq. (\ref{partw})), and $F_0$   is the half-off-shell S-wave scattering BS amplitude presented in the previous section.
Both $\Gamma_i$ and $\Gamma_f$ do not include the external propagators in contrast to the BS amplitude $\Phi$ defined 
in (\ref{bs}) and used to calculate the elastic form factor (\ref{ff}).  

Note that in the approximation given by figure \ref{triangle}, restricted to the one-body electromagnetic current for the interaction of photons with constituents, the current
$\tilde{J}_{\mu}$ is not conserved, that is  $(p'-p)_{\mu}\tilde{J}_{\mu} \neq 0$.
It can be decomposed as:
\begin{equation}\label{ffa}
\tilde{J}_{\mu}=(p_{\mu}+p'_{\mu})F(Q^2)+ (p'_{\mu}-p_{\mu})F'(Q^2)
\end{equation}
In this situation, one can chose one of the two following strategies:
({\it i})~find both $F$ and $F'$ and  use the current (\ref{ffa}), in spite of its non-conservation, to calculate the electro-disintegration cross section;
({\it ii})~start with (\ref{ffa}) and  construct the conserved current
\begin{equation}\label{ffb}
J_{\mu}=\tilde{J}_{\mu}-\frac{q_{\mu}}{q^2}(q\cd\tilde{J})=\left[(p_{\mu}+p'_{\mu})+ \frac{({M_f}^2-M_i^2)}{Q^2}(p'_{\mu}-p_{\mu})\right]F(Q^2)
\end{equation}
which satisfies to $q\cd J=0$ ($q=p'-p$, $q^2=(p'-p)^2$, $Q^2=-q^2$). Due to the constraint $q\cd J=0$, the current $J_{\mu}$ is determined only by the form factor $F(Q^2)$, 
and this is the quantity that will be calculated below. The form factor $F'(Q^2)$,  if necessary, can be found analogously.

It is convenient to carry out the calculations in the system of reference where $p'_0=p_0$ (i.e. $q_0=0$) and $\vec{p}$ and $\vec{p'}$ are collinear (i.e., they are either parallel or anti-parallel to each other, depending on the $Q^2$ value). 
In the elastic case, this system coincides with the Breit frame $\vec{p}+\vec{p'}=0$, where $|\vec{p}|=|\vec{p'}|$ and  $p'_0=p_0$. 
In the inelastic case $M_f \neq M_i$, if $p'_0=p_0$, 
the three-momenta are different $|\vec{p}|\neq|\vec{p'}|$.  

Using this reference frame,  the form factor $F(Q^2)$ can be found by taking the zero component of the current:
$
J_{0}=\tilde{J}_{0}=2p_{0}F(Q^2).
$
That is:
\begin{equation}\label{ffin1}
F(Q^2)=\frac{i}{p_0}\int \frac{d^4k}{(2\pi)^4}\,
\frac{(p_0-k_0)\Gamma_i \left(\frac{1}{2}p -k,p\right)\Gamma_f
\left(\frac{1}{2}p'-k,p'\right)} {(k_0^2-E_{\vec{k}}^2+i\epsilon)[(p_0-k_0)^2-E_{\vec{p}-\vec{k}}^2+i\epsilon]
[(p'_0-k_0)^2-E_{\vec{p'}-\vec{k}}^2+i\epsilon]},
\end{equation}
where  $E_{\vec{k}}=\sqrt{m^2+\vec{k}^2}$  is the on-shell energy  (and similarly for other energies).

This 4D integral  is in practice a 3D one, since the azimuthal integration gives simply a factor $2\pi$.
To calculate it numerically, we treat the singularities in a similar way to what was done for solving the scattering states BS  equation (\ref{BSE}). 
 Namely, we represent each of three propagators as the sum  of the principal value integrals and the $\delta$-function contributions,
that is, symbolically:
$$
F(Q^2)=\int (PV+i\pi\delta)(PV+i\pi\delta)(PV+i\pi\delta)\ldots
$$
The non-zero contributions result from the terms containing two, one and zero $\delta$-functions.
Performing explicit integrations of the $\delta$-functions, we get, correspondingly, the sum of 
1D, 2D and 3D PV integrals. Like in the transformation of the BS equation  from the form (\ref{BSE}) to    (\ref{Eq_F_sym}),  we use the subtraction technique.
In this way we can properly treat the singularities in the integral (\ref{ffin1}) and calculate the transition form factor. Calculating it, we successfully carried out a few tests described in \cite{ck-transit}.
\vspace{0.2cm}
\begin{figure}[h!]
\begin{center}
\begin{minipage}[t]{75mm}
\begin{center}
\includegraphics[width=0.9\textwidth]{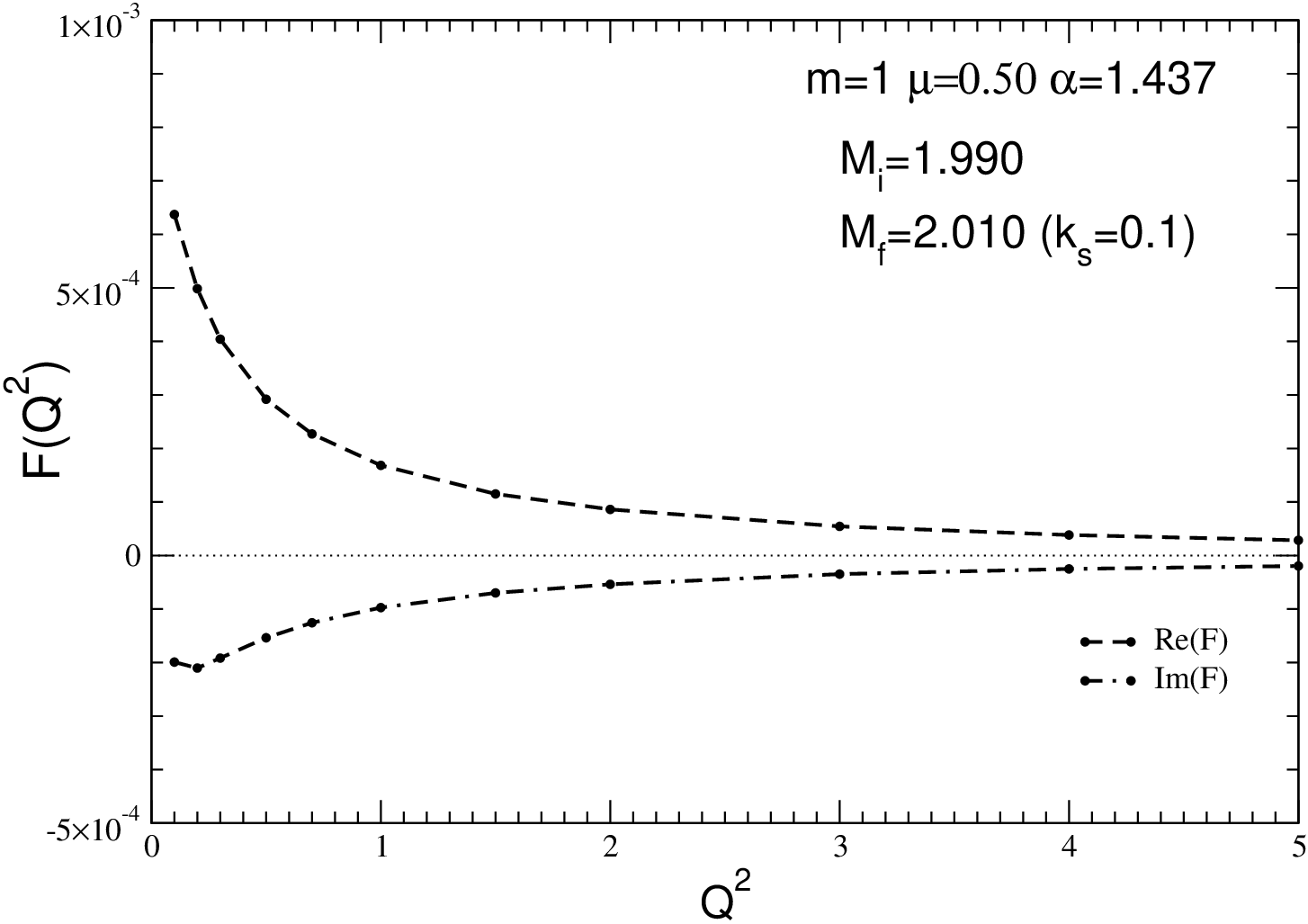}
\caption{Real  (dashed) and imaginary (dot-dashed) parts  of the transition form factor $F(Q^2)$   for $k_s=0.1$.} \label{fig3a}
\end{center}
\end{minipage}
\hspace{0.3cm}
\begin{minipage}[t]{75mm}
\begin{center}
\includegraphics[width=0.9\textwidth]{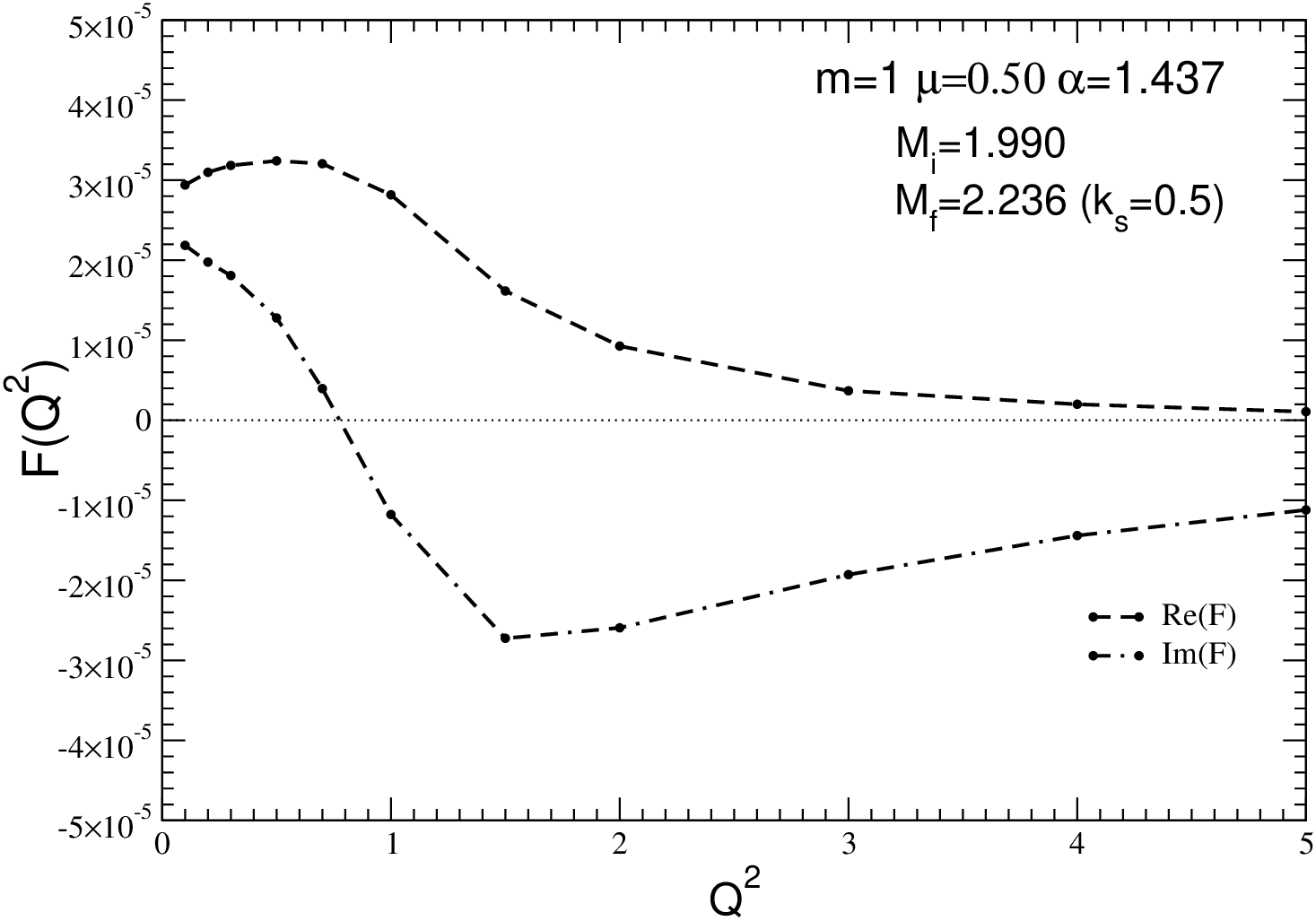}
\caption{The same as in figure \protect{\ref{fig3a}}  for $k_s=0.5$.}\label{fig3b}
\end{center}
\end{minipage}
\end{center}
\end{figure}
\vspace{-0.3cm}

The numerical results are obtained with the bound and scattering state  BS amplitudes found in the one-boson exchange model with the constituent mass $m=1$, exchange boson  mass $\mu=0.5$ and the coupling constant $\alpha=1.437$, providing a bound state with the mass  $M_i=1.99$.
The initial bound state BS vertex function $\Gamma_i$ and the final scattering state amplitude $F_0(k_0,k)$ in $\Gamma_f=16\pi F_0$ have been calculated by the method developed in \cite{KC_LCM_Cracovie_2012,kc_FB20,CK_Baldin_2012,CK_PLB_2013} and described in the previous section.  
Using these solutions, the transition form factor $F(Q^2)$
was calculated by the method presented above. Its real and imaginary parts  (up to a normalization factor) vs. $Q^2$  for the parameters
 $M_i=1.99$ and $M_f=2.01$  ($k_s=0.1$) are shown in figure \ref{fig3a}. 
For these kinematical parameters the transition form factor  is almost real (like the elastic form factor) since the final mass is very close to the initial one.

For $k_s=0.5$ ($M_f=2\sqrt{k_s^2+m^2}=2.236$) and for the same values of other parameters the transition form factor is shown in figure  \ref{fig3b}.
Now, for considerably larger inelasticity (i.e., for larger effective final state mass $M_f$), the imaginary part of form factor is comparable with its real part. 
In our calculations, we have no restrictions for the values of momentum transfer $Q^2$ and the final state mass $M_f$. 

\section{Conclusions}
\label{concl}
The BS equation in the Minkowski space contains (integrable) singularities and therefore its solving requires a special treatment.
We  have described two methods  of finding  solution of the BS equation in Minkowski space. One of them (sect. \ref{sect_nak}) is based on the Nakanishi representation (\ref{bsint}) of the BS amplitude. Other one (sect. \ref{sect_scat}) allows to find the Minkowski solution directly.

Both methods are applied to the bound and scattering states. For the bound states they give the same results. For the scattering states the finding numerical solution by the first method \cite{fsv-2012}  is in progress. By the second method this solution is found \cite{KC_LCM_Cracovie_2012,kc_FB20,CK_Baldin_2012,CK_PLB_2013}: the off-shell amplitude, the scattering length and the phase shifts are calculated. The Minkowski BS amplitude is used to calculate the elastic  \cite{ckm_epja} and transition \cite{ck-transit} (sect. \ref{transit-ff}) e.m. form factors. 

We have mainly discussed the spineless particles. The generalization to the realistic fermion case does not involve any principal difficulties, 
since the singularities of the scalar and fermion propagators are the same. 

The BS approach in Minkowski space, with an appropriate interaction kernel, being applied to the relativistic few-nucleon systems, 
replaces the Schr\"odingier equation and the NN potential used for nonrelativistic description.  
It also makes possible to develop a relativistic version of the Faddeev equations. The off-shell scattering amplitude found 
in this framework (however, full off-shell, not the half one), would be the kernel in the relativistic three-body BS-Faddeev equations.


\end{document}